\documentclass[aps,prl,twocolumn,superscriptaddress,notitlepage]{revtex4-1}

\usepackage{amsmath}
\usepackage{amssymb}
\usepackage{graphicx}
\usepackage{hyperref}

\begin{document}


\title{Structure retrieval at atomic resolution in the presence of multiple scattering of the electron probe}


\author{H. G. Brown}
\affiliation{School of Physics and Astronomy, Monash University, Victoria 3800, Australia}
\author{Z. Chen}
\altaffiliation[Current address: ]{School of Applied and Engineering Physics, Cornell University, Ithaca, New York 14853, USA}
\affiliation{School of Physics and Astronomy, Monash University, Victoria 3800, Australia}
\author{M. Weyland}
\affiliation{Monash Centre for Electron Microscopy, Monash University, Clayton, Victoria 3800, Australia}
\affiliation{Department of Materials Science and Engineering, Monash University, Clayton, Victoria 3800, Australia}
\author{C. Ophus}
\affiliation{National Center for Electron Microscopy, Molecular Foundry, Lawrence Berkeley National Laboratory, California 94720, USA}
\author{J. Ciston}
\affiliation{National Center for Electron Microscopy, Molecular Foundry, Lawrence Berkeley National Laboratory, California 94720, USA}
\author{L. J. Allen}
\affiliation{School of Physics, University of Melbourne, Parkville, Victoria 3010, Australia}	\affiliation{Ernst Ruska-Centre for Microscopy and Spectroscopy with Electrons, Forschungszentrum J\"ulich, 52425 J\"ulich, Germany}
\author{S. D. Findlay}
\altaffiliation[Author to whom correspondence should be addressed: ]{scott.findlay@monash.edu}
\affiliation{School of Physics and Astronomy, Monash University, Victoria 3800, Australia}
%
\date{\today}
\begin{abstract}
The projected electrostatic potential of a thick crystal is reconstructed at atomic-resolution from experimental scanning transmission electron microscopy data recorded using a new generation fast-readout electron camera. This practical and deterministic inversion of the equations encapsulating multiple scattering that were written down by Bethe in 1928 removes the restriction of established methods to ultrathin ($\lesssim 50$ {\AA}) samples. Instruments already coming on-line can overcome the remaining resolution-limiting effects in this method due to finite probe-forming aperture size, spatial incoherence and residual lens aberrations.
\end{abstract}
\maketitle
 Unlike most other types of radiation, electrons interact strongly with condensed matter samples through the Coulomb force. This makes it possible to study very small amounts of material, and modern aberration-corrected transmission electron microscopy (TEM) is capable of sub-50 pm resolution~\cite{erni2009atomic}. However, in all but the thinnest specimens the strong interaction invariably leads to multiple scattering of the probing electrons as expressed in the equations of Bethe~\cite{bethe1928theorie}. Consequently, the recorded images are a non-linear function of the specimen electrostatic potential which describes both the atomistic structure and the electron distribution due to interatomic bonding, complicating the determination of these quantities. Broadly, this problem has been approached in three ways.

The first approach is to assume a structure model and simulate the images by solving the forward scattering problem. Results that agree well with experiment give confidence in the model, whereas significant discrepancies tell against it. So-called directly interpretable imaging modes -- including negative-Cs-imaging in conventional TEM (CTEM)~\cite{jia2004high} and incoherent imaging modes in scanning TEM (STEM)~\cite{pennycook1988chemically} -- aid this approach because despite the multiple scattering the appearance of these images often suggests column locations and constituency in crystals aligned along zone axes. Information on the electron redistribution due to interatomic bonding is not directly measured in this approach but can be predicted via first principles calculations.

The second approach is numerical optimization, solving the forward scattering problem for a sequence of electrostatic potentials refined for successively better fit to the experimental data. A variety of approaches have been proposed~\cite{lentzen1996reconstruction,van2012method,2018arXiv180703886R}, but only a few successfully applied to experimental data \cite{Wang2016LACBED}. Given the multidimensional and nonlinear nature of the problem, optimization approaches often stagnate in local minima. Reducing the dimensionality helps, hence the experimental success of quantitative convergent beam electron diffraction using few-beam conditions to reconstruct the low-order Fourier coefficients of potential most sensitive to bonding~\cite{zuo1999direct,nakashima2011bonding}.

The third approach seeks to solve the inverse scattering problem, to deterministically reconstruct the electrostatic potential from experimental data without at any stage solving the forward problem. Established approaches only achieve this via simplifying assumptions. Multiplicative or phase object approximation reconstructions assume multiple scattering confined to a single plane. This has been achieved at atomic resolution in differential phase contrast~\cite{muller2014atomic} and various forms of ptychography~\cite{putkunz2012atom,yang2016simultaneous,jiang2018electron}, but only yields quantitative results for very thin samples~\cite{Close_2015_CSF,muller2017measurement}. Application to a wider class of samples than ultrathin materials requires solving the inverse scattering problem taking into account multiple scattering.

Spence~\cite{spence1998direct} and Allen et al.~\cite{allen2000inversion} showed that  inversion of multiple scattering in thick samples was possible if the complex-valued scattering matrix could be determined. The original proposal required through-focal-series CTEM images at a carefully controlled series of finely-spaced tilts and has never been experimentally realized. Findlay~\cite{findlay2005quantitative} noted that by reciprocity the required data might be collected in STEM. Recording diffraction patterns at each point in an atomic-resolution raster scan was not then possible, but instrumental developments over the last five years have changed this. In this Letter we describe a procedure for constructing the complex-valued scattering matrix from scanning diffraction data, and invert the multiple scattering to reconstruct the projected electrostatic potential from experimental data obtained from an approximately 230 {\AA} thick Si [110] sample.

\begin{figure}[t]
	\includegraphics[width=\columnwidth]{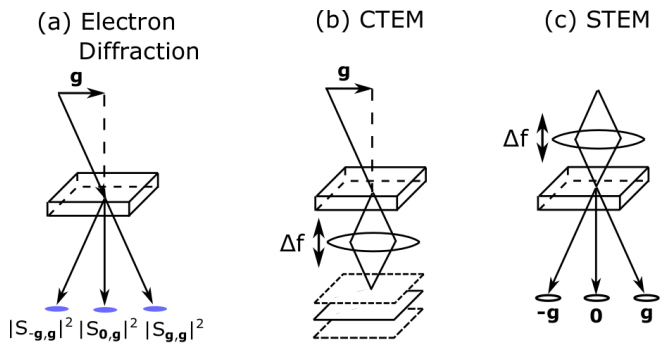}
	\caption{An illustration of different TEM modes: (a) electron diffraction with plane wave illumination tilted to have transverse wave-vector component $\mathbf{g}$; (b) a through-focal series of CTEM images; and (c) diffraction patterns in STEM, which can be recorded as a function of probe position and for different defocus values $\Delta f$.\label{fig:1}}
\end{figure}

We begin with the  Schr\"odinger equation in reciprocal space for a fast electron interacting with the electrostatic potential of a specimen of condensed matter~\cite{van1985image},
\begin{align}
\frac{d \psi_\mathbf{g}(z)}{dz} = -i\pi\lambda g^2\psi_\mathbf{g}(z) + \sum_{\mathbf{h}}i\sigma V_{\mathbf{g-h}}\psi_\mathbf{h}(z)\,.
\label{eq:1}
\end{align}
The $\psi_{\mathbf{g}}(z)$ are Fourier coefficients of the fast electron wave function as a function of depth $z$ in the specimen and Fourier space coordinates $\mathbf{g}$ (magnitude $g$) and $\mathbf{h}$ in the plane perpendicular to the direction of propagation. The Fourier coefficients of the electrostatic potential are denoted by $V_\mathbf{g}$. Inelastic scattering can be described to a good approximation by including an absoprtive component in $V_\mathbf{g}$ \cite{TDSpaper}. The interaction constant $\sigma = 2\pi m_e e\lambda/h^2$, where $m_e$ and $\lambda$ are the (relativistically corrected) mass and wavelength of the electron, $e$ is the electron charge and $h$ is Planck's constant. Equation ($\ref{eq:1}$) is a set of coupled linear equations for which the solution can be written as the matrix-vector product
\begin{align}
\boldsymbol{\psi}(z) = \mathcal{S}(z)\boldsymbol{\psi}(0) \equiv e^{i\mathcal{A}z}\boldsymbol{\psi}(0)\,, \label{eq:2}
\end{align}
where $\boldsymbol{\psi}(z)$ and $\boldsymbol{\psi}(0)$ are vectors containing the Fourier coefficients $\psi_{\mathbf{g}}$ at depth $z$ and the entrance surface, respectively, and the elements of the structure matrix $\mathcal{A}$ are given by $\mathcal{A}_{\mathbf{g,h}} = -\pi\lambda g^2\delta_{\mathbf{gh}}+\sigma V_{\mathbf{g-h}}$, in which $\delta_{\mathbf{gh}}$ is the Kronecker delta. Efficient numerical calculation of Eq.~(\ref{eq:2}) typically proceeds through diagonalisation of the scattering matrix $\mathcal{S}$ (the Bloch wave method~\cite{bethe1928theorie,Humphreys}), or through a split-step evaluation of $\mathcal{S}$, with the operator involving the propagation matrix elements $-\pi\lambda g^2\delta_{\mathbf{gh}}$ and that involving the specimen interaction matrix elements $\sigma V_{\mathbf{g-h}}$ applied in alternating sequence (called the multislice method in the electron microscopy literature~\cite{goodman1974numerical}).

The forward problem essentially consists of calculating ${\mathcal S}$ for a given structure defined by ${\mathcal A}$. In electron diffraction, Fig.~\ref{fig:1}(a), plane wave illumination inclined such that the transverse component of the wave-vector is $\mathbf{g}$ gives diffraction peak intensities proportional to  $|\mathcal{S}_{\mathbf{h},\mathbf{g}}|^2$ at diffraction plane coordinate $\mathbf{h}$. The CTEM image for defocus $\Delta f$ can be calculated from the elements $\mathcal{S}_{\mathbf{h},\mathbf{g}}$ of the scattering matrix via
\begin{align}
I_{\mathrm{CTEM}}(\mathbf{r},\mathbf{g},\Delta f) = \left|\sum_{\mathbf{h}} \mathcal{S}_{\mathbf{h},\mathbf{g}} \exp[2\pi i(h^2\Delta f\lambda+\mathbf{h}\cdot \mathbf{r})]\right|^2\,,
\label{eq:3}
\end{align}
where the summation is over all $\mathbf{h}$ with magnitude less than the probe forming aperture.
The STEM image formed with point detector placed at diffraction coordinate $\mathbf{g}$ can be calculated from the elements $\mathcal{S}_{\mathbf{g},\mathbf{h}}$ of the scattering matrix via
\begin{align}
I_{\mathrm{STEM}}(\mathbf{r},\mathbf{g},\Delta f) = \left|\sum_{\mathbf{h}}\mathcal{S_{\mathbf{g},\mathbf{h}}}\exp[2\pi i(h^2\Delta f\lambda-\mathbf{h}\cdot \mathbf{r})]\right|^2\,,
\label{eq:4}
\end{align}
where again the summation is over all $\mathbf{h}$ with magnitude less than the probe forming aperture.

The inverse problem essentially consists of using measurements on ${\mathcal S}$ to deterministically construct ${\mathcal A}$. Spence~\cite{spence1998direct} and Allen et al.~\cite{allen2000inversion} present a method of achieving this if the complex-valued ${\mathcal S}$ matrix has been determined. Whereas the geometry of Fig.~\ref{fig:1}(a) does not provide sufficient information to determine the phase of the elements of ${\mathcal S}$, the CTEM geometry of Fig.~\ref{fig:1}(b) allows the determination of the amplitude \emph{and} phase of elements $\mathcal{S}_{\mathbf{h},\mathbf{g}}$ corresponding to reciprocal space vectors $\mathbf{h}$ within the imaging aperture by reconstructing the complex-valued exit surface wave from through-focal series for each value of $\mathbf{g}$~\cite{allen2004exit}. Exit wave reconstruction by varying other probe aberrations~\cite{allen2001computational,mcbride2005astigmatic} or specimen tilt~\cite{kirkland1995super} are also possible. The STEM geometry of Fig.~\ref{fig:1}(c) likewise allows the determination of the amplitude \emph{and} phase of elements $\mathcal{S}_{\mathbf{g},\mathbf{h}}$ corresponding to reciprocal space vectors $\mathbf{h}$ within the probe forming aperture from a single through-focal series scan by placing point detectors at points $\mathbf{g}$.

We effect the reconstruction of $V_{\mathbf{g}}$ as a two-part process. First, we reconstruct $\mathcal{S}$ from STEM experiments using a pixelated detector capable of fast read-out ($\gtrsim$100 frames per second) -- instrumentation that has only recently become available~\cite{ophus2014recording} -- such that diffraction patterns can be recorded at each probe position during the STEM raster scan. Second, we invert ${\mathcal S}$ to obtain $\mathcal{A}$, from which the Fourier coefficients $V_\mathbf{g}$ may be read. This is possible because, as follows from Eq.~(\ref{eq:2}), $\mathcal{A}$ and $\mathcal{S}$ share the same eigenvectors, and the eigenvalues of $\mathcal{A}$ can be determined from the eigenvectors of $\mathcal{S}$ using the known diagonals of $\mathcal{A}$ and a number of general symmetries in $\mathcal{A}$ \cite{allen2000inversion}. Since only the eigenvectors of $ \mathcal{S} $ are used as input into the reconstruction whereas information about specimen thickness is contained only within the eigenvalues of $ \mathcal{S} $, an estimate of specimen thickness is not required for the reconstruction -- a particularly advantageous property of this method.

\begin{figure}[h!]
	\includegraphics{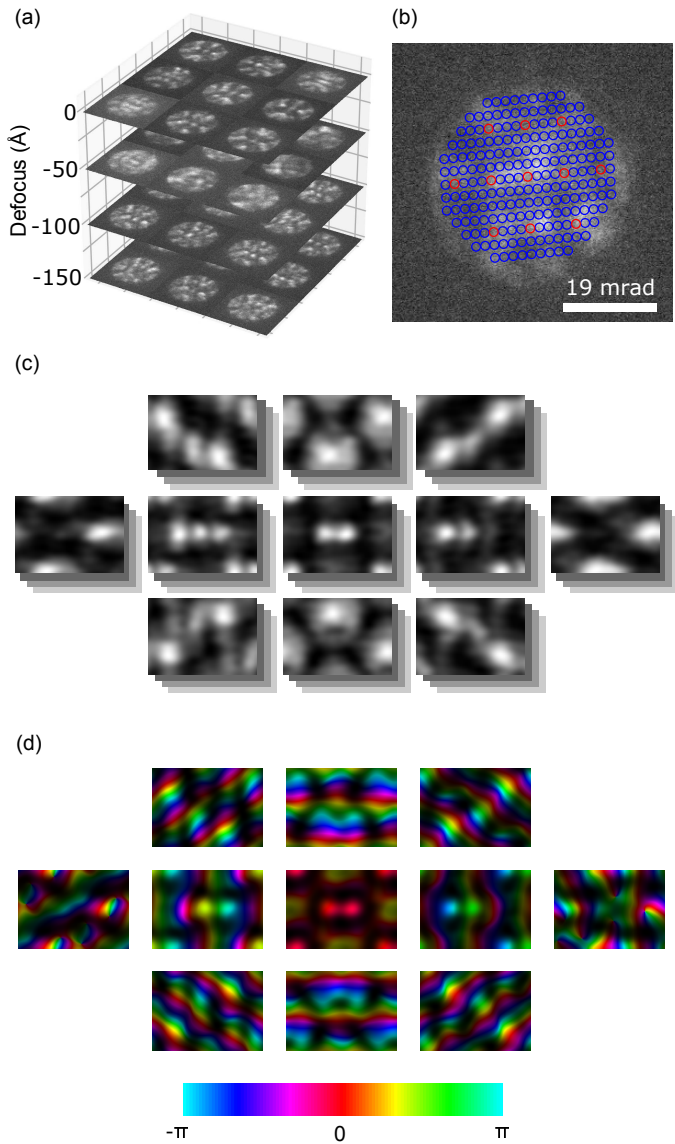}
	\caption{A 4D-STEM scan was recorded at four different defocus values, of which a small subset of the recorded diffraction patterns are shown in (a). A through-focal series of synthesized bright-field STEM images was then constructed using the detectors plotted as circles on the diffraction pattern in (b), the subset formed from the red circles being shown in (c). The complex-valued wave function associated with these through-focal series was then reconstructed via iterative wave function reconstruction~\cite{allen2004exit}, those shown in (d) corresponding to the series shown in (c) (with color denoting the phase). A more complete set of images with their corresponding complex-valued wave function reconstructions is shown in supplementary Figs.~S1 and S2.\label{fig:2}}
\end{figure}

The $\mathcal{S}$-matrix reconstruction process is sketched in Fig.~\ref{fig:2} for experimental data from a [110] oriented silicon crystal recorded on an FEI CETA fast readout electron camera in an aberration-corrected FEI Titan$^3$ 80-300 FEGTEM microscope. The instrument was operated at 305 kV using a 19 mrad probe forming aperture, and data were recorded at four different defocus values ($\Delta f =0$, $-50$, $-100$ and $-150$ \AA)  with a sampling of 0.16 \AA/pixel and a probe dwell time of 3.6 $\mu$s. A snapshot of the diffraction patterns for each scan in the focal series is shown in Fig.~\ref{fig:2}(a). STEM images are formed by plotting, as a function of probe position, the integrated signal in circular regions of the diffraction pattern that are large enough to reduce noise but small enough that the average intensity well-approximates the ideal intensity in a point detector. This is depicted on a representative diffraction pattern in Fig.~\ref{fig:2}(b). A montage of unit-cell-averaged STEM images is shown in Fig.~\ref{fig:2}(c), the $\Delta f = 0$ \AA\ dataset being shown at the top of each stack.

In STEM, the incoherence of the electron source, distortions of the probe scan raster and mechanical vibrations all mean that experimental images exhibit a lower resolution than that expected based on the probe-forming aperture alone. Convolution with an effective source size is typically applied to simulated images to account for these factors. For the experimental data presented here, simulated high-angle annular dark field images needed to be convolved with a Gaussian of full-width at half maximum (FWHM) of 0.86 \AA\ for good agreement with experiment (see Fig. S4), a value consistent with previous measurements on this instrument~\cite{maunders2011practical}. The effective source size needs to be deconvolved from the STEM images, which was done using 10 iterations of the Lucy-Richardson method. Note that this process can only partially ameliorate the effects of spatial incoherence, as Fourier frequencies suppressed below the level of noise cannot be faithfully recovered.

Given the structure of Eq. (\ref{eq:4}), the complex-valued ${\mathcal S}$-matrix row corresponding to the $\mathbf{g}$ coordinate of the STEM detector can be reconstructed using the iterative wave function  retrieval method~\cite{allen2001phase}. A montage of reconstructed complex-valued wave functions from the focal series in Fig.~\ref{fig:2}(c), after deconvolution of effective source size, is shown in Fig.~\ref{fig:2}(d), where color represents the phase. A subtlety of phase reconstruction is that each wave function is only recoverable up to an arbitrary constant phase offset $\Delta \phi_{\mathbf{g}}$, meaning that each of the independently recovered wave functions in Fig.~\ref{fig:2}(d) are potentially offset by these arbitrary phases. This is overcome by using the symmetry $\mathcal{S}_{\mathbf{h},\mathbf{g}}= \mathcal{S}_{-\mathbf{g},-\mathbf{h}}$ \cite{allen2000inversion} and unconstrained minimization to find the optimal set of $\Delta \phi_{\mathbf{g}}$.

\begin{figure}
		\includegraphics[width=0.75\columnwidth]{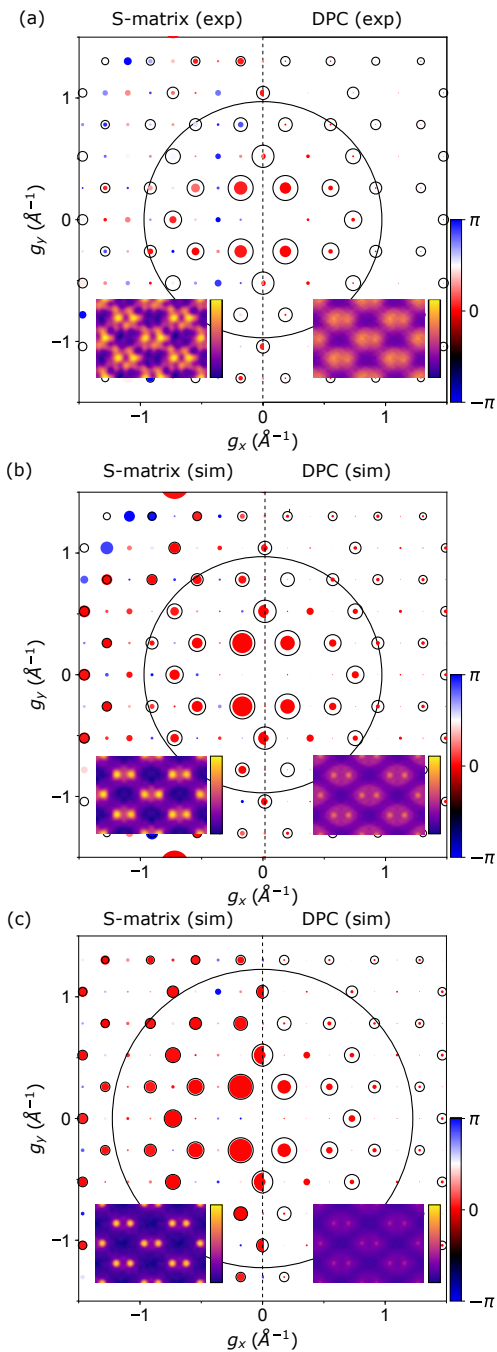}
	\caption{
		(a) An evaluation of the ${\mathcal S}$-matrix and DPC reconstructed potential applied to the 4D-STEM experimental data from Si [110]. The area of each circle/disk in the scatter plot is proportional to $|V_{\mathbf{g}}|$ while the disk color respresents the phase difference between the reconstructed and reference $V_{\mathbf{g}}$, as given by the colorbar. The size of the probe forming aperture is indicated by the large black circle and a real space map of each reconstructed potential is inset for reference. Also shown are ${\mathcal S}$-matrix and DPC reconstructions from simulated data for a hypothetical microscope with improved coherence and aberration correction for (b) 19 and (c) 24 mrad probe forming apertures. Within each sub-figure, the inset real space maps of the reconstructed potential are given on the same absolute scale.
		\label{fig:3}}
\end{figure}

From the same dataset, reconstructions were carried out using both $\mathcal{S}$-matrix inversion and differential phase contrast (DPC) reconstruction~\cite{muller2014atomic} (the latter from the $\Delta f=-100$ \AA\ dataset since mid-plane defocus is optimal for DPC reconstruction~\cite{Close_2015_CSF}). Comparison of the reconstructions is best carried out by comparing the Fourier coefficients of the projected potential $V_{\mathbf{g}}$ (also called the structure factors), with bonding information predominantly residing in the low order Fourier coefficients. Fig.~\ref{fig:3}(a) compares reconstructed Fourier coefficients from the ${\mathcal S}$-matrix inversion, the DPC reconstruction, and simulation. The black circles in Fig.~\ref{fig:3}(a) represent the simulated Fourier coefficients of the projected potential $V_{\mathbf{g}}$, positioned according to the reciprocal space coordinate $\mathbf{g}$, assuming an independent atom approximation. The colored disks represent the reconstructed Fourier coefficients, those on the left half of the figure being reconstructed by inversion of the ${\mathcal S}$-matrix and those on the right half of the figure being reconstructed using DPC. The area of the circles and disks is proportional to the magnitude of the Fourier coefficient, $|V_\mathbf{g}|$. The color of the disks represents the phase \emph{difference} between the reconstructed and simulated Fourier coefficients. Thus a reconstructed Fourier coefficient of potential $V_\mathbf{g}$ is accurately reconstructed (relative to the simulation) if the corresponding disk precisely fills the surrounding circle and is red. 

It is thus seen in Fig.~\ref{fig:3}(a) that the ${\mathcal S}$-matrix inversion gives a better reconstruction of the Fourier coefficients, particularly the amplitudes of the higher order coefficients, although there are some notable deviations among their phases. Both methods appear to give comparably good reconstructions for the lower order Fourier coefficients, though neither is in perfect agreement with simulation.

Since reconstructions are frequently shown as real-space plots of the potential, such plots are inset in Fig.~\ref{fig:3}(a) with the unit cell tiled $2\times 2$ and low-pass filtered to contain only those Fourier coefficients for which ${\bf g}$ is within the probe-forming aperture. The characteristic dumbbell structure of Si [110] is clearly seen. The potential from the ${\mathcal S}$-matrix inversion has better contrast and more finely resolved features than that of the DPC reconstruction, consistent with the larger magnitudes of the high-order reconstructed Fourier coefficients, though the errors in their phases manifest as fine oscillations. That said, we consider the real space version to be somewhat misleading, precisely because errors in high frequency Fourier coefficients tend to have a disproportionally large impact on the visual appearance of the real-space images. For experimental measurement of the electron distribution due to interatomic bonding, it is the low order Fourier coefficients that are of the most interest~\cite{zuo1999direct,nakashima2011bonding}. (Note: interatomic bonding has not been included in the simulations in Fig.~\ref{fig:3}.)

Though for this dataset Fig.~\ref{fig:3}(a) suggests that ${\mathcal S}$-matrix inversion gives only a modest improvement over the DPC reconstruction, the limitations on the ${\mathcal S}$-matrix inversion result from experimental factors whereas the limitations on the DPC reconstruction result from the fundamental break down for thicker samples of its underlying assumption that multiple scattering is confined to a single plane. This is visible in the reconstructions shown in Figs.~\ref{fig:3}(b) and (c), for 19 mrad and 24 mrad convergence angles respectively, performed on data simulated using the $\mu$STEM simulation package~\cite{Allen2015}. Both sub-figures assume aberration-free probes, a spatial incoherence of FWHM 0.4 \AA\ -- achievable on existing microscopes equipped with the latest electron sources~\cite{erni2009atomic, brown2017new} -- and a comparable level of noise to the present experiment.

Much better agreement is seen between the Fourier coefficients from the ${\mathcal S}$-matrix inversion and the simulated reference for all $\mathbf{g}$ within the probe forming aperture in Fig.~\ref{fig:3}(b). Some errors remain in the reconstructed amplitude of the Fourier coefficients due to the so-called truncation effect, the restriction the probe forming aperture places on the size of the $\mathcal{S}$-matrix that can be reconstructed~\cite{findlay2005quantitative}. These errors are reduced by increasing the size of the probe forming aperture, as demonstrated in Fig.~\ref{fig:3}(c) where the Fourier coefficients of potential are more faithfully reconstructed. The remaining small errors are predominantly due to the addition of Poisson noise to the data calculated using $\mu$STEM. We thus conclude that the improved coherence and aberration correction available with the latest generation aberration-corrected electron microscopes will produce improved ${\mathcal S}$-matrix inversion and potential reconstruction in the presence of multiple scattering. 

No similar improvement in the DPC reconstruction is predicted; despite better input data, the DPC reconstructions in Figs.~\ref{fig:3}(b) and (c) are little improved over that in Fig.~\ref{fig:3}(a). This is because, unlike the $\mathcal{S}$-matrix inversion, the DPC reconstruction makes the projection approximation which is poor for this 230 \AA\ thick Si [110] sample.

This Letter has demonstrated a direct inversion from experimental 4D-STEM data to structure in the presence of multiple scattering, thereby overcoming the long-standing limitation of direct reconstruction methods to very thin samples. Newly available fast-readout pixel detector capabilities in an aberration-corrected electron microscope were essential to obtain the necessary scattering dataset. Our results unambiguously show the potential for routine direct solution of the multiple scattering problem given the better source coherence and more precise control of lens aberrations available in instruments currently coming on line.

\begin{acknowledgments}

This research was supported under the Australian Research Council’s Discovery Projects funding scheme (Project No. DP140102538). We acknowledge support from a Network of Excellence grant from Monash University, the use of facilities within the Monash Centre for Electron Microscopy, and the Titan instrument funded under Australian Research Council's LE0454186. Work at the Molecular Foundry was supported by the Office of Science, Office of Basic Energy Sciences, of the U.S. Department of Energy under Contract No. DE-AC02-05CH11231. JC acknowledges additional support from the U.S. Department of Energy Early Career Research Program. The software used to carry out the ${\mathcal S}$-matrix inversion is available on request.

\end{acknowledgments}

%

\end{document}